\documentclass[10pt,conference]{IEEEtran}

\pdfoutput=1
\usepackage[utf8]{inputenc}
\usepackage{fontawesome5}

\newcommand{\jia}{$\text{\faMale}_A$}
\newcommand{\lasse}{$\text{\faMale}_{PM}$}
\newcommand{\Aio}{$\text{\faMale}_{IO}$}

\newcommand{\Asm}{$\text{\faMale}_{SM}$}

\usepackage[T1]{fontenc}
\usepackage{url}
\usepackage{flushend}
\usepackage{xspace}
\usepackage[skip=5pt]{caption}
\usepackage[inline]{enumitem}
\usepackage{ifthen}
\usepackage{balance}
\usepackage[hidelinks]{hyperref}
\usepackage[numbers]{natbib}

\usepackage{microtype}
\usepackage{multirow}
\usepackage[]{siunitx}
\newcommand*\np[2][z]{
\ifx z#1%
$\num{#2}$%
\else%
$\qty{#2}{#1}$%
\fi\xspace%
}

\usepackage{fp}
\usepackage{xfp}

\usepackage{graphicx}
\usepackage{subcaption}
\usepackage{pifont}

\usepackage{booktabs}
\usepackage{multirow}
\usepackage{threeparttable}
\usepackage{xcolor}
\usepackage{array,makecell}
\usepackage{diagbox}
\usepackage{tabularx}

\newcommand{\foo}{\makebox[0pt]{\textbullet}\hskip-0.5pt\vrule width 1pt\hspace{\labelsep}}

\usepackage{mathtools}
\usepackage{amsmath}

\usepackage{tcolorbox}
\usepackage{pgfplots}
\usepackage{pgfplotstable}

\usepackage{enumitem}
\usepackage[framemethod=TikZ]{mdframed}
\usepackage[scaled]{beramono} 

\usepgfplotslibrary{statistics}
\usetikzlibrary{calc,patterns,fpu} 
\pgfplotsset{compat=1.14} 

\pgfplotsset{
    /pgf/declare function={
        Floor(\x) = round(\x-0.49);
    },
    show sum on top/.style={
        /pgfplots/scatter/@post marker code/.append code={%
            \path let \p1=($(normalized axis cs:%
                        \pgfkeysvalueof{/data point/x},%
                        \pgfkeysvalueof{/data point/y})%
                        -(normalized axis cs:\pgfkeysvalueof{/data point/x},0)$)
            in node[
                at={(normalized axis cs:%
                        \pgfkeysvalueof{/data point/x},%
                        \pgfkeysvalueof{/data point/y})%
                },
                anchor={-90*sign(\y1)},yshift={sign(\y1)*2pt}
            ]
            {\pgfmathprintnumber[fixed, precision=1]{\pgfkeysvalueof{/data point/y}}};
        },
    }
}

\definecolor{blau_1a}{RGB}{93,133,195}
\definecolor{blau_2a}{RGB}{0,156,218}
\definecolor{gruen_3a}{RGB}{80,182,149}
\definecolor{gruen_4a}{RGB}{175,204,80}
\definecolor{gruen_5a}{RGB}{221,223,72}
\definecolor{orange_6a}{RGB}{255,224,92}
\definecolor{orange_7a}{RGB}{248,186,60}
\definecolor{rot_8a}{RGB}{238,122,52}
\definecolor{rot_9a}{RGB}{233,80,62}
\definecolor{lila_10a}{RGB}{201,48,142}
\definecolor{lila_11a}{RGB}{128,69,151}

\definecolor{blau_1b}{RGB}{0,90,169}
\definecolor{blau_2b}{RGB}{0,131,204}
\definecolor{gruen_3b}{RGB}{0,157,129}
\definecolor{gruen_4b}{RGB}{153,192,0}
\definecolor{gruen_5b}{RGB}{201,212,0}
\definecolor{orange_6b}{RGB}{253,202,0}
\definecolor{orange_7b}{RGB}{245,163,0}
\definecolor{rot_8b}{RGB}{236,101,0}
\definecolor{rot_9b}{RGB}{230,0,26}
\definecolor{lila_10b}{RGB}{166,0,132}
\definecolor{lila_11b}{RGB}{114,16,133}

\newboolean{showcomments}
\setboolean{showcomments}{true}

\newcommand{\ShowAbsoluteNumber}[1]{%
\ifnum #1<10%
{\hspace*{0pt}#1}%
\else%
\ifnum #1<100%
{\hspace*{0pt}#1}%
\else%
\ifnum #1<1000%
{\hspace*{0pt}#1}%
\else%
{\numprint{#1}}%
\fi%
\fi%
\fi%
}

\newcommand{\ShowPercentage}[2]{%
\FPeval\percentage{round(#1/#2*100,0)}%
\FPeval\percentageOneDecimal{round(#1/#2*100,1)}%
\ifnum \percentage=0%
{\np[\%]{\FPprint{percentageOneDecimal}}}%
\else%
\ifnum \percentage<10%
{\np[\%]{\FPprint{percentageOneDecimal}}}%
\else%
{\np[\%]{\FPprint{percentageOneDecimal}}}%
\fi%
\fi%
\xspace
}
\newlength\BARSIZE  
\newcommand{\inlinechart}[2]{%
\FPeval{\BLACKBARSIZE}{#1/#2}\textcolor{black!80}{\rule{\BLACKBARSIZE\BARSIZE}{1.6ex}}%
\FPeval{\BLACKBARSIZE}{1 - (#1/#2)}\textcolor{black!10}{\rule{\BLACKBARSIZE\BARSIZE}{1.6ex}}%
}

\newcommand*\percent[3][v]{%
\ifx q#1%
    \np{#2}/\np{#3}(\ShowPercentage{#2}{#3})\else%
\ifx s#1%
    \ShowPercentage{#2}{#3}\else%
\ifx p#1%
    \np{#2}(\ShowPercentage{#2}{#3})\else%
\ifx c#1%
    \inlinechart{#2}{#3}%
\else%
    \np{#2}%
    \ifx r#1%
        /\np{#3}%
    \fi%
    \hspace*{0.5ex}(\ShowPercentage{#2}{#3}) %
    \inlinechart{#2}{#3}%
    \xspace
\fi\fi\fi\fi%
}

\definecolor{mygray}{RGB}{240,240,240}

\newcommand{\answer}[2]{\vspace{.2cm}{\centering\setlength{\fboxrule}{0.1pt}\fbox{\colorbox{mygray}{\parbox{0.95\columnwidth}{\textbf{Answer to RQ#1}. #2}}}\vspace{.2cm}}}

\DeclareRobustCommand{\step}[1]{\tikz[baseline=(char.base)]{\node[shape=circle,text=black,draw=black, fill=white,inner sep=0.2pt] (char) {#1};}}

\ifthenelse{\boolean{showcomments}}
 { }
        { }

\definecolor{eminence}{RGB}{108,48,130}
\definecolor{weborange}{RGB}{255,165,0}
\definecolor{frenchplum}{RGB}{129,20,82}
\definecolor{darkgreen}{RGB}{10, 92, 10}

\mdfdefinestyle{mpdframe}{
    nobreak                     =true,
    backgroundcolor             =black!3,
    frametitlebackgroundcolor   =black!15,
    frametitlerule              =false,
    roundcorner                 =5pt,
    innermargin                 =0.3cm,
    innerleftmargin             =0.3cm,
    innerrightmargin            =0.3cm,
    innertopmargin              =0.3cm,
    innerbottommargin           =0.3cm,
    shadowsize                  =1pt
}

\usepackage{cleveref}

\usepackage[frozencache, cachedir=minted-cache]{minted}
\setminted{breaklines,breakanywhere}
\setmintedinline{breaklines,breakafter=.}

\title{Wolves in the Repository: A Software Engineering Analysis of the XZ Utils Supply Chain Attack}

\author{
    \IEEEauthorblockN{Piotr Przymus}
    \IEEEauthorblockA{
        Nicolaus Copernicus University in Toru\'n, Poland \\
        Email: piotr.przymus@mat.umk.pl,\\
        ORCID: 0000-0001-9548-2388}
    \and
    \IEEEauthorblockN{Thomas Durieux}
    \IEEEauthorblockA{
        TU Delft \& Endor Labs, The Netherlands\\
        Email: thomas@durieux.me\\
        ORCID: 0000-0002-1996-6134
        }
}

\begin{document}

\maketitle

\begin{abstract}
The digital economy runs on Open Source Software (OSS), with an estimated 90\% of modern applications containing open-source components. While this widespread adoption has revolutionized software development, it has also created critical security vulnerabilities, particularly in essential but under-resourced projects. This paper examines a sophisticated attack on the XZ Utils project (CVE-2024-3094), where attackers exploited not just code, but the entire open-source development process to inject a backdoor into a fundamental Linux compression library. Our analysis reveals a new breed of supply chain attack that manipulates software engineering practices themselves -- from community management to CI/CD configurations -- to establish legitimacy and maintain long-term control. Through a comprehensive examination of GitHub events and development artifacts, we reconstruct the attack timeline, analyze the evolution of attacker tactics. Our findings demonstrate how attackers leveraged seemingly beneficial contributions to project infrastructure and maintenance to bypass traditional security measures. This work extends beyond traditional security analysis by examining how software engineering practices themselves can be weaponized, offering insights for protecting the open-source ecosystem.
\end{abstract}

\begin{IEEEkeywords}
security, project management
\end{IEEEkeywords}

\section{Introduction}

In today's digital infrastructure, Open Source Software (OSS) forms the bedrock of modern technology, powering everything from critical infrastructure to everyday applications. Studies estimate that 90\% of modern applications contain open source components~\cite{lenarduzzi2020open}, making OSS security not just a software engineering concern, but a matter of global digital security. While OSS offers numerous advantages including cost efficiency, flexibility, and transparency, it also means that vulnerabilities can have cascading effects across the entire software ecosystem.
This critical dependence on OSS creates a paradoxical situation: some of the most widely-used software components are maintained by small teams or even single individuals working on a voluntary basis~\cite{dias2018drives}. Unlike proprietary software with dedicated security teams and resources, these essential OSS projects often rely on unpaid maintainers who balance project responsibilities with full-time jobs and personal commitments~\cite{10.1145/3106237.3106246}.

The risks of this model have been demonstrated repeatedly through high-profile security incidents. The Heartbleed bug in OpenSSL affected millions of servers, while the Log4Shell vulnerability sent shockwaves through the industry, both highlighting the fragility of critical open source infrastructure. These incidents underscore a significant concern: the prevalence of important OSS projects maintained by small teams of volunteer developers who often experience burnout and lack motivation but continue out of a sense of duty~\cite{oneillInternetRunsFree, tulili2023burnout, lifshitz2021digital}, supplemented only by sporadic external contributions.

This creates a risky situation with wide-reaching effects.
\begin{itemize}[leftmargin=1em]
    \item The ``bus factor'' – a measure of risk associated with information and capabilities being concentrated in a small number of individuals – becomes alarmingly low.
    \item Maintainers face pressure, shouldering the responsibility for projects that may be critical to numerous dependent systems.
    \item The broader community becomes vulnerable to disruptions if the maintainer becomes unavailable or unable to continue their work. A notable example is the case of the core-js library, which boasts 32 million weekly downloads on NPM~\cite{corejs}.
    \item Security risks escalate, ranging from potential backdoors and crypto mining to the injection of vulnerabilities into the source code.
\end{itemize}

The recent XZ Utils attack (\href{https://nvd.nist.gov/vuln/detail/CVE-2024-3094}{CVE-2024-3094}) represents a particularly sophisticated evolution of these threats. In this incident, malicious actors methodically gained control of the XZ Utils project, a critical component in many Linux systems, with the explicit intention of injecting a vulnerability. This attack targeted the OpenSSH server, potentially exposing numerous servers to exploitation. The vulnerability allowed attackers to either send arbitrary payloads via SSH before authentication or bypass password authentication entirely. The attack is described in detail in \Cref{sec:attack}.

What makes this attack particularly concerning from a software engineering perspective is the attackers' exploitation of software development practices themselves. While previous security incidents often focused on code-level vulnerabilities, the XZ Utils attack demonstrates how attackers can manipulate the entire development process. They not only introduced malicious code but also engaged in various software engineering practices to establish credibility and maintain control over the project. This included contributing to community management, modifying CI configurations, updating the build system, and even moving the project to GitHub – actions that, on the surface, appeared to be beneficial to the project's development~\cite{dias2021makes}.

As software systems increasingly rely on OSS components, it becomes imperative to study not just the technical vulnerabilities but also how attackers can exploit software engineering practices themselves. This is particularly critical as the increasing automation in software engineering could simplify these complex and costly attacks in the near future. Through a detailed analysis of the XZ Utils attack, we aim to answer the following research questions:

\begin{itemize}[leftmargin=2.5em]
    \item[RQ1] What is the timeline of the XZ Utils attack?
    \item[RQ2] What software engineering practices have been used by the attackers?
    \item[RQ3] What are the consequences and impacts of the attack?
\end{itemize}

The findings from this analysis have implications beyond just security. They speak to fundamental questions about trust, governance, and sustainability in open source communities. Through this analysis, we seek to contribute to the ongoing dialogue on enhancing the security and sustainability of open-source software. Our study provides a comprehensive dataset of the attack timeline, offers insights into the software engineering practices employed by attackers, and proposes direction for identifying potential problematic contribution in OSS projects as well as highlighting in inerrant complexity of the task.

The contribution of this paper lies in providing a comprehensive perspective of the attack timeline, examining the software engineering practices exploited, and analyzing their cascading impact on the open-source community. 

The OSS community has thoroughly investigated the attack's technical aspects, but a centralized, systematic analysis is still needed. Unlike reports that focus on specific elements of the attack, our study offers a holistic view, integrating qualitative and quantitative methods to validate findings across all available data rather than relying on selective examples.

Through understanding these attack mechanisms, we aim to equip the software engineering community with actionable insights to enhance the resilience of open-source ecosystems.

\subsection{Methodology}
This study uses a mixed-methods approach that combines qualitative and quantitative analyses to investigate the software engineering practices involved in the XZ Utils attack. Our analysis relies on a comprehensive dataset constructed from multiple sources, including the project's Git repository, GitHub event logs, mailing list archives, and external security databases. We systematically examined commit histories, contributor activity, and development artifacts to reconstruct the attack timeline and assess the attacker's involvement in the project.

To contextualize our findings, we reviewed existing reports about the attack, many of which focused on isolated aspects like backdoor implementation, contributor behavior, or security implications. 
Based on this initial analysis, we refined our methodology to provide a holistic perspective, addressing gaps in prior analyses and validating findings with the entire dataset rather than selective examples.

To classify the attack activities, we employed an open coding methodology, as there was no established framework available for analyzing software supply chain attacks of this nature. We manually annotated all recorded interactions, including code reviews, emails, and discussions, to develop a comprehensive understanding of the attacker’s role within the community. Additionally, we utilized automated annotation techniques~\cite{piotr_przymus_patchscope_2024} to analyze code changes at a granular level, identifying the specific types of modifications introduced by the attacker. This categorization process involved iterative discussions among the authors to refine classifications and resolve ambiguities.

We conducted statistical analysis to validate hypotheses related to the involvement of multiple accounts, examining patterns of activity across contributors.

Given the unique nature of this attack, we recognize that our analysis is based on a single case study. However, the potential severity of this incident underscores the importance of deriving generalizable insights to enhance security practices in open-source software development.

\section{Background}
In this background section, we present the XZ Utils attacks, the key actors, and how this vulnerability has been discovered. 
This provides a general understanding of the technical and social aspects that were required to perform this attack.

\subsection{Key Actors}

The \texttt{XZ Utils} attack involved several significant actors, each playing a crucial role in the incident's unfolding\footnote{Since our work focuses on the impact of the attack on software engineering practices rather than specific individuals, we have anonymized the actors involved.}:

\begin{itemize}[leftmargin=1em]

    \item \textbf{Primary Maintainer ($\text{\faMale}_{PM}$)}:  The longstanding maintainer of \texttt{XZ Utils} since 2009. The Primary Maintainer's diligent upkeep likely protected the project from earlier security threats. Their long-standing commitment to the project underscores the challenges faced by solo maintainers of critical open-source infrastructure.

    \item \textbf{Attacker (\jia)}: The principal persona behind the attack on \texttt{XZ Utils}. The Attacker became involved in the project's maintenance in 2021, gradually establishing a position of trust within the community before executing the attack.

    \item \textbf{Anomalous Committer ($\text{\faMale}_{AC}$)}: A contributor who introduced the \texttt{ifunc} resolver to \textit{XZ Utils}, a crucial functionality that facilitated the attack. The Anomalous Committer's involvement raises important questions. Their contributions required specific expertise but were limited in number. Their GitHub activity ceased abruptly after these contributions.
    To this day, it remains unclear whether these patches were genuine improvements or part of the broader attack strategy.

    \item \textbf{Incident Observer (\Aio) }: A PostgreSQL developer at Microsoft who discovered and investigated the vulnerability. The Incident Observer's role in uncovering the threat highlights the importance of cross-project vigilance in the open-source community. The discovery process is detailed in \Cref{sec:discovery}.

    \item \textbf{Additional Actors ($\text{\faMale}_{AA}$)}: The attack involved several other personas, likely created to add credibility and pressure. Puppet accounts on mailing lists were used to advocate for various purposes related to the attack.\footnote{19 May 2022: \url{https://www.mail-archive.com/xz-devel@tukaani.org/msg00562.html}, 07 Jun 2022: \url{https://www.mail-archive.com/xz-devel@tukaani.org/msg00566.html}, \url{https://www.mail-archive.com/search?l=xz-devel@tukaani.org&q=from:\%22Dennis+Ens\%22}} 
    Multiple GitHub accounts with minimal activity were created. These accounts introduced attack-related code changes before becoming inactive, a pattern that retrospectively appears suspicious. 

    \item \textbf{Secondary Maintainer ($\text{\faMale}_{SM}$)}: A long-term contributor to the Gentoo Linux distribution and XZ Utils. After the attack, the Secondary Maintainer stepped up and became the second main maintainer.
\end{itemize}

The diversity and complexity of malicious actors' roles demonstrate the sophisticated nature of the attack, highlighting the need for enhanced scrutiny of contributor patterns and project governance in open-source software.

\subsection{XZ Utils Vulnerability Description}\label{sec:attack}

\textit{XZ Utils} attack (\href{https://nvd.nist.gov/vuln/detail/CVE-2024-3094}{CVE-2024-3094}) involved interactions within several critical software components essential to modern Linux systems: \texttt{OpenSSH} and \texttt{systemd}. 
\texttt{OpenSSH} is a widely used tool for remote accessing servers and embedded systems, if compromised it would open the door to many critical systems~\cite{6838249}. 
\texttt{systemd} is a service manager for Linux, designed to start, stop, and manage system processes. 
\texttt{OpenSSH} is targeted for getting access to computers while \texttt{systemd} was required to be able to compromise \texttt{OpenSSH}.

XZ Utils project was probably targeted because it is a dependency of \texttt{systemd} and could therefore be used to inject the vulnerability inside \texttt{OpenSSH} using  \texttt{systemd} privileges.
Additionally, XZ Utils was appealing to attackers due to its status as a low-traffic repository managed by a single developer with a small community (around 10 active members on the project's IRC channel).

More precisely, the malicious backdoor in XZ Utils is designed to inject code into the OpenSSH server (sshd) on the target machine. This enables specific remote attackers, who have a designated private key, to accomplish one of the following:

\begin{enumerate}
\item Send arbitrary payloads via SSH, which are executed before the authentication step, or
\item Bypass SSH password authentication, allowing the attacker to authenticate with any password.
\end{enumerate}

The payload is introduced into the OpenSSH server through \texttt{liblzma} (part of XZ Utils), which contains the malicious code. Interestingly, OpenSSH itself does not depend on \texttt{liblzma}, but this dependency is indirectly enforced by Linux distributions that patch OpenSSH to support \texttt{systemd}. When a service starts, it should notify \texttt{systemd} that it has started. The preferred method is to use the function \texttt{sd\_notify} implemented in \texttt{libsystemd}. 
OpenSSH is patched and linked against \texttt{libsystemd} to support \texttt{sd\_notify}.

However, one piece was still missing in XZ Utils for code injection: the ifunc feature, enabling dynamic selection of function implementations at runtime. This optimizes software for various hardware without recompiling and was introduced to liblzma by the commit \href{https://github.com/tukaani-project/xz/commit/23b5c36fb71904bfbe16bb20f976da38dadf6c3b}{419f55f} on June, 22 2023, motivated by a small performance boost. 

At this stage, \textit{XZ Utils} with \texttt{ifunc} and being loaded by \texttt{systemd} is able to replace system functions. The \texttt{ifunc} resolver was run early to force function selection, adding to the attack's sophistication.
The attacker replaced the \texttt{RSA\_public\_decrypt} function, originally intended for validating RSA signatures. 
The attacker could therefore bypass the \texttt{OpenSSH} authentication and connect to any \texttt{OpenSSH} server.
More precisely, the malicious code examines the RSA public modulus (\texttt{N} value) provided within the RSA structure (the fourth argument of \texttt{RSA\_public\_decrypt}). This modulus is fully controlled by the connecting SSH client, i.e., the attackers.

The attack appears to specifically target amd64 systems running glibc on Debian or Red Hat derived distributions, although other systems may also be vulnerable. This vulnerability was patched by a build system targeting these specified Linux distros. Notably, the build patch was obfuscated; it included a shell-based extractor for the payload code, delivered into the project as a binary test case— a common practice for testing compression libraries. 
The build system patch was not in the Git version of the code but was included in the distribution package generated for each new software release. It's normal for the release package to differ from the Git version due to the tooling used to simplify the repackaging process.

\subsection{Discovery of the Vulnerability}\label{sec:discovery}

The \textit{XZ Utils} vulnerability was uncovered by \Aio. His discovery, aided by a stroke of luck, began with his observation of anomalous behavior: unusually high CPU usage by sshd processes during SSH logins. This led him to investigate further, ultimately revealing the attack, on March 28, 2024,.

\Aio traced the malicious code to apparently legitimate updates committed by \jia. The backdoor affected XZ Utils versions 5.6.0 and 5.6.1, but only in the official archived download packages (``tarballs'') and not the Git distributions. This discrepancy highlighted the attackers' careful approach to evading detection.

In his disclosure, \Aio  noted: ``Given the sustained activity over several weeks, it seems likely the committer was directly involved or their system was severely compromised.'' He expressed skepticism about the latter scenario, citing the maintainer's public communications about the purported fixes.

\section{SE Practices of the Attacker}

In this section, we present our empirical analysis of the XZ Utils attack, examining both its chronology and the software engineering practices employed. Our investigation addresses the following research questions:
\begin{itemize}
\item[RQ1] \textbf{What is the timeline of the XZ Utils attack?}
This research question examines the attack's chronological progression, identifying critical events that facilitated its execution. We analyze the temporal patterns of both the attacker's activities and the original maintainer's contributions to understand how the attack unfolded.
\item[RQ2] \textbf{What software engineering practices have been used by the attackers?}
This research question investigates the software engineering methodologies and techniques deployed by the attackers during their project maintenance period. We examine how these practices were utilized to advance their malicious objectives while maintaining project credibility.
\item[RQ3] \textbf{What are the consequences and impacts of the attack?}
This research question evaluates the broader implications of the attack. While early detection prevented widespread exploitation, we analyze both the immediate effects on \textit{XZ Utils} and the broader ramifications for the OSS ecosystem's security and governance practices.
\end{itemize}

\subsection{Dataset}
To conduct a comprehensive analysis of the XZ Utils attack, we assembled a multifaceted dataset that captures both quantitative project metrics and qualitative contextual information. Our data collection methodology was designed to provide a complete picture of the attack's evolution, encompassing activities both within and outside the GitHub ecosystem.
The dataset comprises four primary components:
\begin{enumerate}
\item \textbf{Git Repository Analysis}: Detailed commit history metrics, including quantitative analysis of code changes categorized by type (source code, documentation, test cases, and translations), obtained through automatic line annotation of each contribution using~\cite{piotr_przymus_patchscope_2024}.

\item \textbf{GitHub Project Metrics}: Comprehensive project activity data, including issue tracking, pull request history, and GitHub event data sourced from GitHub Archive~\cite{gharchive}\footnote{\url{https://docs.github.com/en/rest/using-the-rest-api/github-event-types}}. These metrics provide insights into project governance and community interactions.
\item \textbf{XZ Utils Mailing List Archives}: Historical development communications predating the GitHub migration, encompassing crucial decision-making discussions and technical debates from both the mailing list and SourceForge discussions.
\item \textbf{Security Context Analysis}: Integration of external security data from Have I Been Pwned~\cite{hunt2019have} to evaluate potential compromise vectors of contributor accounts and analyze email address usage patterns across different services.
\end{enumerate}

We have made this dataset publicly available to facilitate further investigation of this sophisticated OSS attack. 
\Cref{tab:dataset} summarizes the key components.
This dataset provides insights into both the attack's progression and its impact on the project's technical and social dimensions. Beyond collecting raw data from diverse sources, we conducted detailed manual analyses to extract and annotate significant patterns. This included a systematic examination of \jia early contributions and code reviews from the mailing list, as well as a thorough commit-by-commit analysis to identify relevant software engineering tools, practices, and patterns.

\begin{table}[htbp]
\centering
\caption{Overview of Dataset Components}
\label{tab:dataset}
\begin{tabularx}{0.4\textwidth}{@{}Xr@{}}
\toprule
\textbf{Data Source} & \textbf{\#} \\
\midrule
Commits & 1020 \\ 
Annotated lines & 91263\\
GitHub Events & 2944 \\
Mailing List Messages & 307\\
Unique emails (checked in 'Have I Been Pwned') &  62 \\ 
\bottomrule
\end{tabularx}
\end{table}

\subsection{RQ1. Timeline}

In this research question, we analyze the chronological progression of the XZ Utils attack, examining the relative contributions of both the attacker and the primary maintainer. Our analysis quantifies individual engagement through multiple metrics, including commit patterns and user interactions. Additionally, we evaluate community growth trends as external factors that may influence core developer dynamics, potentially accelerating trust accumulation and role transitions within the project team.

\begin{figure*}[t] 
    \centering
    \includegraphics[width=1.0\textwidth, trim={0cm 0cm 0cm 0cm}, clip]{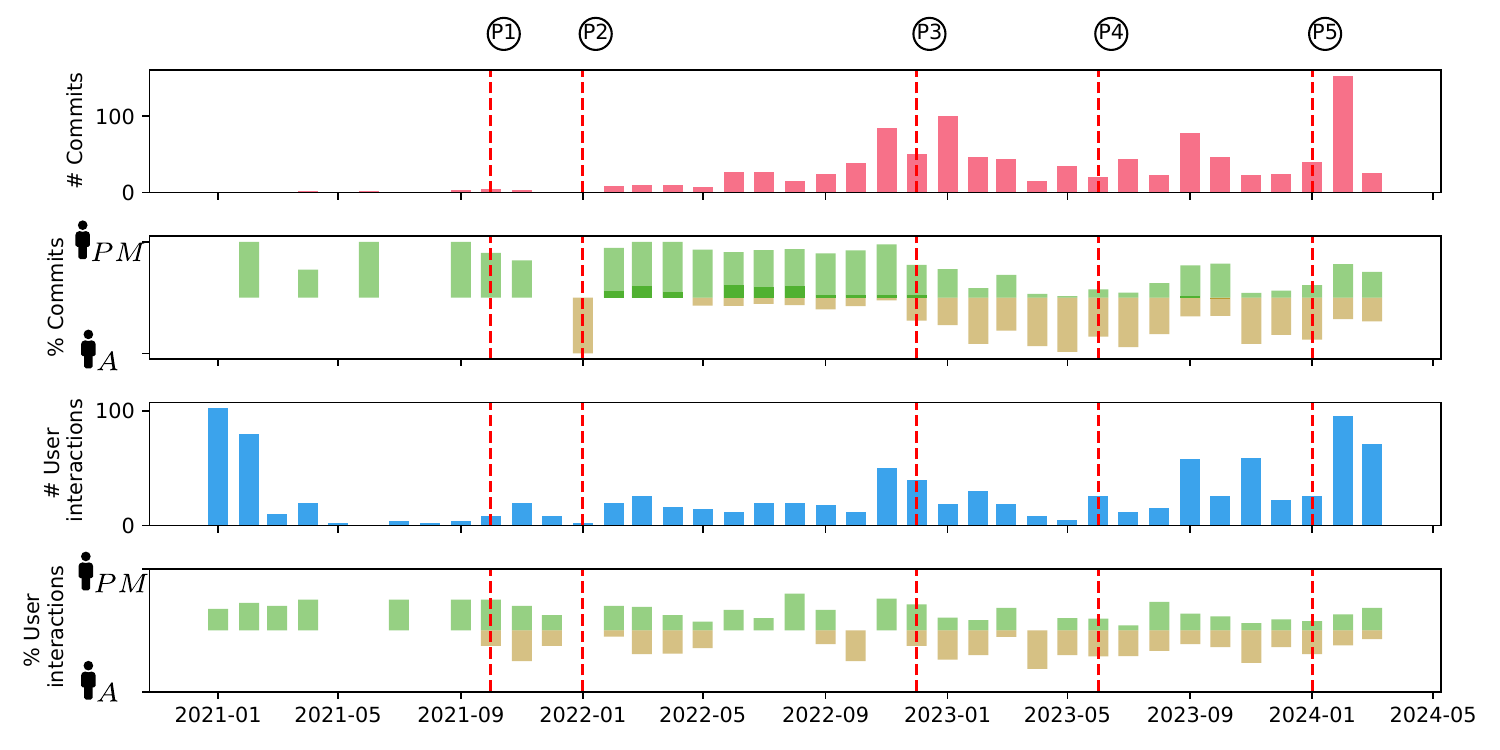}
    \caption{\jia and \lasse actions, aggregated monthly and divided into commits and user interaction events. Vertical lines indicate the start of a phase with major events described in \Cref{tab:timeline}. The $\#$ plots display the number of events of each type, while the $\%$ plots show the percentage of events. If the changes are a result of collaboration between \jia and \lasse, they are marked with a darker color.
    }
    \label{fig:timeline-attacks}
\end{figure*}

\Cref{tab:timeline} presents a comprehensive chronology of \jia activities within the project. We have identified and documented all events significant to understanding both the attack's execution and its contextual environment. \Cref{fig:timeline-attacks} illustrates the temporal distribution of commits and activities between \jia and the primary maintainer. Metrics labeled with `\#' represent absolute counts of specific activity types within each month, while `\%' indicators show the relative proportion of activities performed by either \jia or \lasse. Activities are categorized into commits and user interactions, where commits represent direct code contributions, and collaborative commits (involving both \jia and \lasse) are distinguished using darker shading. User interactions encompass all community engagement, including mailing list participation, SourceForge discussions, and GitHub-based activities such as issue discussions and code reviews.

We proposed dividing the attack activities into distinct phases (see~\Cref{tab:timeline}), denoted by vertical lines in \Cref{fig:timeline-attacks}. Each phase begins with a new type of activity by the attacker in the project (\step{P0}-\step{P3}) or starts with a commit introducing functionality needed for the attack (\step{P4}-\step{P5}). 

\begin{table}[t]
\setlength{\tabcolsep}{3pt}
\caption{Timeline of the \textit{XZ Utils} attack}\label{tab:timeline}
\begin{tabularx}{0.495\textwidth}{@{\,}cr <{\hskip 1pt} !{\foo} >{\raggedright\arraybackslash}X@{}}
\toprule
       \multirow{3}{*}{\step{P1}}  & 2021.10.29 & \jia 1st patch proposal (\texttt{.editorconfig} file) \href{https://www.mail-archive.com/xz-devel@tukaani.org/msg00512.html}{[link]} \\
       & 2021.11.10 & \jia 2nd patch proposal (changes the test framework) \href{https://www.mail-archive.com/xz-devel@tukaani.org/msg00514.html}{[link]} \\
       & 2021.11.12 & \jia 1st code review \href{https://www.mail-archive.com/xz-devel@tukaani.org/msg00518.html}{[link]} \\
\midrule

\multirow{4}{*}{\step{P2}} & 2022.01.21 & \jia 1st authored commit (sha 6468f7e41)\\
       & 2022.02.22 & \jia helped with the 1st commit accepted to repository\\
       & 2022.05.19 & Pressure for more maintainers \href{https://www.mail-archive.com/xz-devel@tukaani.org/msg00562.html}{[link]} \\
       & 2022.06.07 & Additional pressure for more maintainers \href{https://www.mail-archive.com/xz-devel@tukaani.org/msg00566.html}{[link]} \\
\midrule

\multirow{5}{*}{\step{P3}} & 2022.12.12 & \jia created the GitHub organization and the GitHub repository for \textit{XZ Utils} \\
       & 2022.12.12 & \jia set his email as the default contact \href{https://github.com/tukaani-project/.github/commit/3358b3d9ad3d47774194fef9129a479adaa4a4fc}{[link]} \\
       & 2022.12.13 & First release made by \jia \\
       & 2023.01.11 & \jia started making announcements on the mailing list \href{https://www.mail-archive.com/xz-devel@tukaani.org/msg00634.html}{[link]} \\
\midrule

\multirow{3}{*}{\step{P4}} & 2023.06.22 & Introduction of \texttt{ifunc} \\
       & 2023.10.12 & CRC code usng \texttt{ifunc}\\
       & 2024.01.19 & \jia migrated the website to the GitHub page \\
       \midrule
       
\multirow{8}{*}{\step{P5}} & 2024.01.22 & First backdoor commit\\
       & 2024.02.24 & XZ 5.6.0 is released \\
       & 2024.02.26 & Commit in CMakeLists.txt sabotages the Landlock security feature. \\
       & 2024.03.04 & The backdoor leads to issues with Valgrind \\
       & 2024.03.09 & Two ``test files'' are updated, CRC functions are modified, and the Valgrind issue is ``fixed'' \\
       & 2024.03.09 & XZ 5.6.1 is released \\
       \midrule

       \multirow{12}{*}{\step{A}}& 2024.03.28 & Bug is discovered, Debian and RedHat notified \\
       & 2024.03.28 & GitHub accounts of both \lasse and \jia   are suspended. \lasse account was reinstated on 2024-04-02 \\
        & 2024.03.28 & Debian rolls back XZ 5.6.1 to the 5.4.5-0.2 version \\
       & 2024.03.29 & Bug published on the OSS-security mailing list \\
       & 2024.03.29 & RedHat confirms the XZ backdoor was shipped in Fedora Rawhide and Fedora Linux 40 beta \\
       & 2024.03.30 & Debian shuts down builds and starts the process of rebuilding \\
       & 2024.04.02 & XZ's main developer recognizes the backdoor incident \\
       & 2024.04.30 & RedHat releases RHEL 9.4 \\
       & 2024.06.29 & Debian releases 12.6 \\
\bottomrule
\end{tabularx}
\end{table}

The first main observation is that \jia was involved from 2021.10.29 to 2024.03.28, spanning 2.6 years. Although all malicious attacks occurred in 2024, some suspicious activities were observed earlier. For instance, \jia used his email address for bug reports and try replaced the test framework with his library. 
He has been active since his first post on the mailing list, as highlighted in the timeline and commits.
However, we can observe that the type of activity changed over the years. 
In 2022, \jia focused on making code contributions, while in 2023, he performed most maintenance tasks, such as handling translations and maintaining the CI/CD and build system. 
The coding activity resumed in the early months of 2024 in preparation for the release targeted for RedHat 9.4 at the end of April.

Interestingly, \jia had \texttt{write} permission on the Git repository since 2022, but it does not appear that these permissions were exploited in the attack. He did not push malicious code nor rewrite the project's history using \texttt{push --force}. However, he did use his position as a maintainer for the attack, as described in \Cref{sec:attack}. The malicious code was not deployed in the GitHub release but was instead deployed during the manual release made by \jia, specifically targeting RedHat and Debian. He waited until a month before the official RedHat release and close to Debian release to push his attack into the repository during a period of high project activity, as shown in \Cref{fig:timeline-attacks}.

We now explore the five proposed phases.
Before phase \step{P1}, the project was in a stagnation period, with little development taking place. Although the mailing list was active at the beginning of 2021, it slowed as we approached the starting point of \step{P1}.
During phase \step{P1} \jia has established his presence by proposing some entry-level contributions to the project. Later on, he is involved in code review (doing a reasonable job even with fairly complicated code reviews). 

The project dynamics changed considerably starting from phase \step{P2}, when \jia first commit was accepted. During this period, there was an overall increase in the number of commits and communications, with \jia responsible for some of them, either directly or by co-authoring with \lasse. Notably, we observed eight new users on the mailing list, marking a 27.5\% increase from the previously active 29 users (six of those users are only active during this period ). Additionally, there was a rise in the number of commits containing ``Thanks'' messages, with 65 such commits recorded, the highest number compared to any previous year of the project (previous years had a median of 16, a maximum of 49, and a total of 255, with peak activity in 2009-2010). This period saw the highest traffic of such commits since the project's inception. Additionally, two new users on the mailing list pressured the main maintainer to increase efforts. 
During this period, \jia slowly gained trust and became more involved in the project.

The next phase, \step{P3}, started when the project transitioned to the GitHub platform. \jia initiated this process by setting up the organization on GitHub. During this time, \jia became the primary contributor and took over maintainership, engaging more with user interactions and announcing new releases. The project migrated to GitHub, changing the previous development model. The mailing list was less used, while GitHub issues, pull requests, and reviews became the norm. During this period, 17 users actively engaged through events such as pushing code, submitting and reviewing pull requests, raising and commenting on issues. The majority of these interacting accounts have an extended history on GitHub (13 accounts created before 2021, with extended history), lending credibility to their contributions. This shift to GitHub attracted new experienced developers contributions to the project.

Phase \step{P4} starts with the introduction of the \textit{ifunc} resolver. As mentioned, this could be a genuine contribution and/or part of the attack. The facts are that this functionality was introduced by a developer with low activity on GitHub, who created their account after 2021 and did not contribute anything after the attack was discovered. \jia is still responsible for the majority of commits and user interaction during this period. During this phase, we observed interactions from 24 developers, of which four were also present in the previous phase. The majority of these users (19) created their accounts before 2021 and have credible histories (with one exception: a user who created an issue that motivated move of website to GitHub \href{https://github.com/tukaani-project/xz/issues/78}{[link]}.).

The last stage of the attack, \step{P5}, started with the first malicious commit by \jia. During this stage, we observe increased efforts leading to new version releases. The efforts are led by \jia and \lasse, who helped with many commits during this period. Some changes appear to happen to the development process, such as changing the way translations are handled (\jia modified this process, which \lasse opposed). During this stage, we observe 19 unique active developers, with 5 of them contributing in the previous phases. The majority of these developers created their accounts before 2021 and have credible histories.

After \step{P5}, we observe a massive increase in developers interested in the project. We observed 322 users who actively interacted with the project, with only 5 of them being active in \step{P5}. This can be compared to the recent Log4j vulnerability, which also resulted in increased interest in the development of the project.

Retrospective analysis of the timeline (\Cref{tab:timeline}) and commit patterns (\Cref{fig:timeline-attacks}) reveals how \jia systematically established his position as a project maintainer through multiple vectors:
\begin{enumerate*}
    \item Sustained long-term contribution patterns
    \item Strategic engagement in code review processes
    \item Strategic advocacy for additional maintainers during 2022
    \item Infrastructure modernization through GitHub migration, securing organizational control
    \item Gradual assumption of community management responsibilities
\end{enumerate*}

The challenge in detecting such attacks lies in their legitimate appearance---only eight commits contained malicious code, while the majority of activities aligned with standard open-source development practices. While \jia comprehensive involvement across multiple project aspects might appear unusual in retrospect, distinguishing such patterns from enthusiastic legitimate contributors remains a significant challenge, particularly given the diverse software engineering practices employed (detailed in RQ2~\ref{sec:rq2}).

\answer{1}{The XZ Utils attack spanned 2.6 years (2021-2024) and followed a phased strategy: trust-building through mailing list contributions, gradually taking on maintainer duties, modernizing infrastructure via GitHub migration, and finally injecting malicious code. Though the actual attack involved only eight commits in 2024, the attacker built credibility through consistent project involvement, including code reviews and community management. The timing was aligned with upcoming RedHat and Debian releases, coinciding with peak project activity. Other accounts displayed similar focused engagement patterns, though no direct links were found. This timeline shows how routine project involvement can be methodically exploited, underscoring the difficulty of distinguishing between well-intentioned and malicious contributors.}

\subsection{RQ2. Software Engineering Activities Employed by the Attacker}\label{sec:rq2}
To address our second research question, we analyze the software engineering practices employed throughout the attacker's engagement with the XZ Utils project. Our investigation reveals a sophisticated strategy that leveraged modern development practices not only to establish credibility and gain control but ultimately to execute the attack.

During our analysis of \jia contributions, we identified a diverse portfolio of software engineering practices spanning multiple years. \Cref{tab:se-practices} presents a comprehensive overview of these practices, their apparent purposes, and their security implications within the context of the attack. Our main observation is that \jia was involved in all types of activity. 

\begin{table*}[t]
\centering
\caption{Software Engineering Practices Employed by the Attacker.}
\label{tab:se-practices}
\begin{tabularx}{\textwidth}{@{}lXlr@{}}
\toprule
\textbf{Practice} & \textbf{Apparent Purpose} & \textbf{Security Implication} & \textbf{Examples}\\
\midrule
Community Management & Build trust and influence & Increased authority in decision-making & \href{https://www.mail-archive.com/xz-devel@tukaani.org/msg00553.html}{[link]}\\
Setup CI/CD & Modernize infrastructure & Control over automated processes and change contact email. & \href{https://github.com/tukaani-project/xz/commit/9f05c27a58ce8cd7803079aa295e41c24665ce6e}{[link]}\\
Core Code Contributions & Address genuine issues & Establish non-threatening presence and introduce exploitable features& \href{https://github.com/tukaani-project/xz/commit/72d2933bfae514e0dbb123488e9f1eb7cf64175f}{[link]} \\
Code Review Participation & Show collaborative spirit & Establish non-threatening presence &  \href{https://www.mail-archive.com/xz-devel@tukaani.org/msg00518.html}{[link]} \\
Translation & Demonstrate engagement & Establish non-threatening presence \\
Build System Changes & Improve build & Establish non-threatening presence & \href{https://github.com/tukaani-project/xz/commit/bb740e3b117f1a3c65152d01e5755523a908ecb1}{[link]}\\
Test Expansion & Improve code quality & Hide malicious payloads & \href{https://github.com/tukaani-project/xz/commit/e2870db5be1503e6a489fc3d47daf950d6f62723}{[link]}\\
GitHub Migration & Enhance project visibility and reduce git.tukaani.org traffic & Gain ownership of the project and organization & \href{https://www.mail-archive.com/xz-devel@tukaani.org/msg00634.html}{[link]}\\
Website Migration & Simplify Edition & Change contact email & \href{https://www.mail-archive.com/xz-devel@tukaani.org/msg00642.html}{[link]}\\
Mailing List Engagement & Demonstrate expertise & Influence community perception & \href{https://www.mail-archive.com/xz-devel@tukaani.org/msg00593.html}{[link]}\\
\bottomrule
\end{tabularx}
\end{table*}
\begin{figure}[t]
    \centering
    \includegraphics[width=0.48\textwidth]{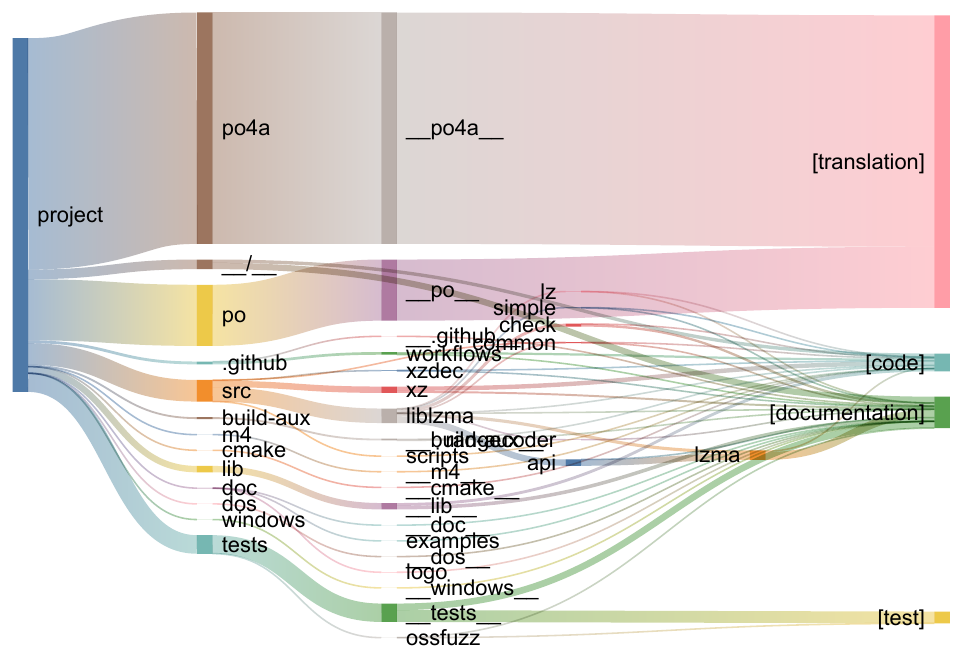}
    \caption{\textbf{\jia Contributions by Project Directory and Type.}  
This plot visualizes all \jia Git commits, with each line automatically annotated and aggregated. The flow represents the total number of modified lines from commits. Nodes on the left show \textbf{project directories}, while the right categorizes contributions (\textit{code, documentation, tests, translations}). The \textbf{width of connections} indicates the volume of changes, highlighting where \jia focused most efforts.}

    \label{fig:jia-contributions}
\end{figure}

Interestingly, the main contribution of \jia during those years was not in the source code but instead in the documentation and translation (see \Cref{fig:jia-contributions}).
Those types of contributions required less involvement and they still contribute to a regular activity to the repository and therefore contribute to his credibility.
\jia even replaced the existing translation process with his own translations. 
The original process used external resources to do it while \jia seems to use automatic translation tools or copy-pasting translating from other projects without probably knowing the languages (hypothesis due to the number of targetted languages).

In addition, he was de facto community manager of the project. 
He was handling the announcements and releases and answered questions and issues on GitHub.
He was more active in this regard than the main contributor.
Those activities also contribute to credibility and presence in the project.

The migration activities put in place by \jia have a more direct malicious impact.
When \textit{XZ Utils} migrated to GitHub, \jia created the organization and the repository and therefore gained ownership of them.
He also used this opportunity to create an issue template where his email was used instead of the email of the main contributor.
This also happened when the website migrated from a self-hosting to the GitHub page.

Many of the presented activities were the responsibility of the main maintainer. 
\lasse managed documentation, community management, code review, release, and translation, and over time, those activities were mostly transferred to \jia (see \Cref{fig:timeline-attacks} and \Cref{fig:jia-contributions}).

The usage of automation facilitated those transitions. 
During our review of the commits, we identified a set of tools and practices that \jia has used. 
In particular, we observe in commits that he aimed at fixing \texttt{cmake} warning, he also used a grammar checker to identify typos in the code and documentation as well as translation software.
Those tools simplified and helped \jia identify potential contributions.
Indeed, the translation and documentation fixes are the most common contributions by \jia as presented in \Cref{fig:jia-contributions}, those contributions mostly happened during 2022 and 2023, see \Cref{fig:timeline-attacks}.

Today, with the dominance of Large Language Models (LLM) for automatizing software engineering tasks, it is likely that even more automatization could be done at a lower cost for the attacker. This specific point is discussed in \Cref{sec:automatization}.

\answer{2}{Our analysis reveals how the attacker methodically employed a comprehensive range of software engineering practices, with documentation and translation activities surprisingly dominating over code contributions. This type of activity enables low-risk engagement that built credibility while avoiding intense scrutiny. The attacker systematically assumed control through multiple vectors: community management (becoming the de facto project spokesperson), infrastructure modernization (gaining control through GitHub migration and CI/CD setup), and strategic automation of routine tasks. Notably, they gradually displaced the main maintainer's responsibilities across documentation, community management, code review, and translations, sometimes using automation tools to facilitate these transitions. The attack demonstrates a sophisticated understanding of open-source dynamics, where seemingly beneficial contributions—particularly non-code activities—can be weaponized to establish authority and create attack vectors while minimizing detection risk.}

\subsection{RQ3. Impact of the Attack on the XZ Utils Project and Broader OSS Community}\label{sec:rq3}

Our final research question examines the immediate and potential long-term impacts of the XZ Utils attack on both the project itself and the wider open-source software community. 
This incident has prompted broad discussion and a reassessment of practices in the OSS community. 
Table \ref{tab:community-impact} summarizes key areas of impact and emerging practices. These areas will be explored in more detail later in the text.

\begin{table}[tbh]
\centering
\caption{Impact and Emerging Practices in the OSS Community}
\label{tab:community-impact}
\begin{tabularx}{0.48\textwidth}{@{}lX@{}}
\toprule
\textbf{Practice Area} & \textbf{Implementation Strategies} \\
\midrule
Contributor Vetting & Enhanced background checks for new maintainers \\
Dependency Management & Increased scrutiny of transitive dependencies \\
Build Process Security & Emphasis on reproducible builds and verifiable build artifacts \\
Governance Models & Move towards multi-stakeholder governance for critical projects \\
Funding and Support & Increased focus on sustainable funding for critical OSS infrastructure \\
Security Tooling & Development of new tools for detecting anomalous contributor behavior \\
Education and Training & Enhanced security training for OSS maintainers and contributors \\
\bottomrule
\end{tabularx}
\end{table}

The first obvious consequence is the attack itself. Fortunately, the malicious release did not progress beyond the testing phase, limiting its impact on the community. To the best of our knowledge, the backdoor was not exploited. The community reacted promptly by reverting the release, and GitHub banned \jia and (temporarily) \lasse, to prevent further damage.

A second consequence was the necessity to review almost three years of commits to ensure no malicious intent remained. This substantial effort required \lasse to scrutinize all contributions from \jia. In total, eight commits were confirmed to be malicious. Some other commits were reverted to reduce complex maintenance (for example all commits related to \texttt{ifunc} functionality). This process led to a thorough analysis that enhanced the project's quality and security.

Paradoxically, the attack has led to some positive outcomes. XZ Utils is now more active than it was before \jia involvement. The project has gained attention, attracting new contributors and a long term contributor step up as second maintainer \Asm.

The attack has sparked various initiatives and discussions within the broader OSS community. Several projects are working to reduce their dependencies, potentially as a response to this incident. CMake was considering hardening \texttt{check\_c\_source\_compiles}, while Systemd is reducing dependencies. GNU binutils is removing dependency on libjansson, and OpenSSH is exploring ways to remove the need for downstream patches for libsystemd.

In terms of build system improvements, Autoconf is addressing bugs and adding syntax checks to code snippets. Groff is distributing bootstrap and bootstrap.conf. There's also a movement towards converting from autotools to Makefiles, as evidenced by esr's ``autodafe'' tool.

New security tools have emerged in response to the incident. The development of ``distro-backdoor-scanner'' for auditing Linux distribution packages and the introduction of ``backseat-signed'', a new supply-chain security tool (available at \url{https://github.com/kpcyrd/backseat-signed}), demonstrate the community's proactive approach to preventing similar attacks in the future.

The XZ Utils incident highlighted key areas for the OSS community: the importance of regular code reviews, active contributors, and stronger security in project management. The community's quick response helped limit damage, while increased scrutiny improved project health and security. Overall, it raised awareness of supply chain risks, encouraging more proactive maintenance and security practices.

Moving forward, the OSS community should implement stricter verification for critical projects, using improved tools and practices to detect and prevent supply chain attacks. Encouraging a culture of security awareness and proactive measures among contributors and maintainers will be key to avoiding future incidents.

\answer{3}{The XZ Utils attack has had broad, lasting effects on the open-source ecosystem. Though contained before exploitation, it led to a thorough three-year review uncovering eight malicious commits. Ironically, the incident has strengthened the project, increasing activity, bringing in new contributors. The broader OSS community has responded with concrete actions: major projects like CMake and Systemd are reducing dependencies; new security tools, such as distro-backdoor-scanner and backseat-signed, have been developed; and build systems are being fortified, evidenced by Autoconf updates and tools like autodafe. These changes reflect a shift toward better contributor vetting, secure build processes, collaborative governance, and sustainable funding for critical OSS infrastructure.}

\section{Discussion}

The XZ Utils attack presents a complex case study that illuminates several critical issues in open-source software (OSS) security and governance. Our analysis reveals the sophisticated nature of the attack and its implications for the broader OSS ecosystem.

\subsection{Sophistication and Patience in Modern OSS Attacks}

The scope and duration of the attacker's involvement in the XZ project highlight a concerning trend in OSS security threats. The attacker's three-year investment in building credibility and gradually assuming control demonstrates a level of patience and sophistication that challenges traditional security models. This approach allowed the attacker to bypass many of the community's implicit trust mechanisms, raising questions about the effectiveness of current contributor vetting processes in long-term scenarios.  Given the sophistication and extended involvement, detecting such attacks can be highly challenging in both open-source projects and commercial environments.

The lucky discovery by \Aio underscores the potential for substantial damage had the attack gone undetected. This near-miss emphasizes the need for more robust, proactive detection mechanisms that can identify anomalous behavior patterns over extended periods.

\subsection{Challenges in Attack Prevention and Detection}

While post-analysis reveals some potentially suspicious behaviors by the attacker (e.g., rapid changes in expertise levels, disproportionate time investment), these actions are often indistinguishable from those of enthusiastic, skilled contributors. The dilemma faced by maintainers like \lasse is clear: on one hand, rejecting or over-scrutinizing seemingly beneficial contributions risks alienating valuable community members and slowing project progress; on the other, accepting them without sufficient scrutiny could leave the project vulnerable to security risks.

\lasse diligent review process, including release files, likely delayed the attacker's progress. However, this level of vigilance is unsustainable for most projects and maintainers. This reality points to the need for automated tools and community structures that can share the burden of security oversight without stifling legitimate contributions.

\subsection{The Double-Edged Sword of Automation in SE}\label{sec:automatization}

The rise of automation tools, including LLMs and various software engineering techniques, presents both opportunities and risks for OSS security. While these tools can enhance productivity and quality, they also have the potential to be exploited for malicious purposes, such as generating believable contributions to build false credibility.

To mitigate these risks while preserving the benefits of automation, we propose the following strategies:

\begin{enumerate}
    \item \textbf{Attribution of Automated Contributions:} Crediting automated changes to tools rather than individuals (similar to Dependabot) could prevent the accumulation of unearned credibility.
    
    \item \textbf{Trusted and Signed Automatic Contributions:} Developing a framework for verifying and trusting automated contributions could help distinguish between legitimate automation and potential attack vectors.
    
    \item \textbf{Enhanced Analysis of Contribution Patterns:} Implementing systems to analyze the evolution of a contributor's expertise and engagement over time could help identify unusual patterns that warrant closer inspection.
\end{enumerate}

\subsection{Balancing Openness with Security}

The XZ Utils case highlights the tension between open-source collaboration and robust security. The community’s quick response and strategy discussions show a commitment to addressing these issues, though balancing stricter security with open-source principles remains delicate.

\subsection{Long-term Implications for OSS Development}

This incident may catalyze significant changes in OSS development practices:

\begin{enumerate}
    \item \textbf{Governance Models:} A shift towards multi-stakeholder governance for critical projects could distribute responsibilities and reduce single points of failure.
    
    \item \textbf{Contributor Evaluation:} More nuanced systems for evaluating long-term contributor intentions and impacts may emerge, potentially leveraging AI and machine learning for pattern recognition.
    
    \item \textbf{Education and Training:} Enhanced security awareness training for maintainers and contributors could become standard practice in the OSS community.
    
    \item \textbf{Funding and Support:} Increased recognition of the critical nature of certain OSS projects may lead to more sustainable funding models and support structures.

\item \textbf{Project Development Behavior Evaluation:} Observing changes in development behavior can be a useful indicator for assessing various risks associated with a project, including security, long-term support, and overall stability.

\end{enumerate}

\subsection{Future Research Directions}

This case study opens several future research directions:

\begin{enumerate}
    \item Developing more sophisticated threat models specific to long-term OSS infiltration attempts.
    \item Creating adaptive security measures that evolve with attacker techniques.
    \item Investigating the psychological aspects of maintaining high-stakes OSS projects to better support maintainer well-being and resilience.
    \item Exploring the ethical implications and potential regulations surrounding the use of AI in OSS contribution.
\end{enumerate}

In conclusion, while the XZ Utils attack exposed vulnerabilities in the OSS model, it also serves as a catalyst for innovation in OSS security and governance. By learning from this incident and implementing thoughtful safeguards, the software engineering community can work towards an open-source ecosystem that maintains its collaborative spirit while significantly enhancing its security and sustainability.

\section{Related Work}

The security of open-source software (OSS) has been a topic of ongoing research and debate in the software engineering community. Payne \cite{payne2002security} provided an early examination of OSS security, highlighting both the potential benefits of the "many eyes" approach to finding vulnerabilities and the risks associated with publicly available source code. This dual nature of OSS security continues to be a central theme in more recent research.

The vulnerability fixing process in OSS projects has been investigated by Canfora et al. \cite{canfora2020investigating}, who identified peculiarities and challenges specific to the open-source model. Their work underscores the importance of understanding the unique dynamics of OSS development when addressing security concerns. Wang et al. \cite{wang2019detecting} further explored this area by studying the "0-day" vulnerabilities in OSS, revealing the challenges of secret security patch management in an open environment.

The issue of trust in OSS has been examined from various angles. Hissam et al. \cite{hissam2002trust} discussed the relationship between trust and vulnerability in open-source software, highlighting the complexities of relying on community-driven development for critical systems. More recently, Guizani et al. \cite{guizani2021long} explored the challenges of contributing to large OSS organizations, providing insights into the social and technical barriers that can impact project security and sustainability.

The security risks specific to package ecosystems have gained increased attention in recent years. Zimmermann et al. \cite{zimmermann2019small} conducted a comprehensive study of security threats in the npm ecosystem, revealing the "small world with high risks" nature of widely-used JavaScript packages. Building on this work, Scalco et al. \cite{scalco2022feasibility} investigated the feasibility of detecting malicious injections in npm packages, proposing new approaches to identify potential security threats.

The broader implications of OSS security on software engineering practices have also been explored. Mouratidis et al. \cite{mouratidis2005security} presented a case study on modeling secure information systems, emphasizing the need to integrate security considerations throughout the software development lifecycle. This work highlights the importance of proactive security measures in both open-source and proprietary software development.

More recently, Zhou et al. \cite{zhou2024oss} conducted an analysis of malicious packages in OSS ecosystems "in the wild," providing valuable insights into real-world attack patterns and vulnerabilities. Their work contributes to a growing body of research aimed at understanding and mitigating security risks in the OSS landscape.

The incident involving the core-js library, as reported by The Register \cite{corejs}, serves as a stark reminder of the potential vulnerabilities in the OSS ecosystem, particularly when critical projects rely heavily on individual maintainers. This case highlights the need for robust governance models and support systems for key OSS infrastructure.
Our work builds upon this foundation of OSS security research, focusing specifically on the software engineering practices employed in sophisticated, long-term attacks on high-impact open-source projects. By analyzing the XZ Utils incident in detail, we aim to contribute new insights into the evolving landscape of OSS security and the challenges faced by the open-source community in maintaining the integrity of critical software infrastructure.

\section{Threats to Validity}

We acknowledge several threats to the validity of our study, categorize them into internal and external threats, and discuss our mitigation strategies where applicable.
\subsection{Internal Threats to Validity}
Our analysis primarily relies on publicly available data, lacking access to private conversations and complete IRC channel history. This limitation may result in incomplete insights into the attacker's interactions and decision-making processes. To mitigate this, we cross-referenced available data sources and used multiple indicators to support our findings.

Attribution of contributions poses another challenge, as some commits may have been prepared jointly but committed by a single developer. We addressed this by analyzing commit messages, pull request discussions, and issue comments to provide additional context, though some ambiguity may remain.
Researcher bias, particularly hindsight bias in interpreting the attacker's motivations and strategies, is a concern. We based our analysis on observable actions and artifacts, with multiple researchers independently reviewing the data and interpretations to reduce individual bias.

\subsection{External Threats to Validity}
The unique characteristics of the XZ Utils project, such as its critical role in many systems and small maintainer team, may limit the generalizability of our findings to projects with different attributes. We attempted to identify general principles and patterns applicable to a broader range of projects while acknowledging the specific combination of factors in this case.
Our analysis is based on a single, sophisticated attack, and attackers may adapt their tactics in response to increased awareness and security measures. To address this, we focused on underlying principles and patterns rather than specific tactics where possible.
The high-profile nature of the XZ Utils project may have influenced the open-source community's response to this attack. The observed changes in practices may not be uniformly adopted across all open-source projects. We distinguished between responses likely to be widely adopted and those that may be specific to high-impact projects.
Finally, our study's limited temporal scope may not fully capture long-term impacts and changes in the open-source ecosystem. We acknowledge the need for longitudinal studies to fully understand the long-term implications of this attack on open-source software development practices.

While we have taken steps to address these threats where possible, we recognize their potential impact on the generalizability and interpretation of our findings. Future research involving a broader range of case studies and longer-term analysis could help to further validate and extend our conclusions.

\section{Conclusion}

The XZ Utils attack (CVE-2024-3094) serves as a stark reminder of the vulnerabilities inherent in the open-source software ecosystem, particularly for high-impact projects maintained by small teams or individual developers. Our analysis of this sophisticated attack reveals several key insights:

\begin{enumerate}
    \item Attackers employed a wide range of software engineering practices, including community management, CI configuration updates, and infrastructure modernization, to establish credibility and maintain long-term control over the project.
    \item The evolution of attacker activities over time demonstrated a careful, patient approach that mimicked legitimate contributor patterns while gradually introducing malicious elements.
    \item Automation tools and modern development practices were leveraged not only to improve the project but also to obfuscate malicious intentions, highlighting the dual-edged nature of these technologies in OSS.
\end{enumerate}

These findings underscore the need for a multifaceted approach to enhancing OSS security:

\begin{itemize}
    \item Improved community vigilance and peer review processes, especially for projects with high downstream impact.
    \item Development of sophisticated tools for detecting anomalous contribution patterns and potential security risks.
    \item Greater support and resources for critical OSS projects, reducing the burden on individual maintainers and lowering the risk of burnout or compromise.
    \item Enhanced education and guidelines for OSS maintainers and contributors on recognizing and mitigating potential security threats.
\end{itemize}

The aftermath of the XZ Utils attack has already spurred significant changes in the OSS community, including more rigorous review processes, efforts to reduce unnecessary dependencies, and the development of new security auditing tools. However, these reactive measures must be complemented by proactive strategies to build a more resilient OSS ecosystem.

Future research should focus on developing more sophisticated threat models for OSS projects, creating adaptive security measures that can evolve with attacker techniques, and exploring sustainable funding and support models for critical open-source infrastructure. Additionally, the psychological aspects of maintaining high-stakes OSS projects deserve further investigation to better support the mental health and resilience of key contributors.

In conclusion, while the XZ Utils attack exposed serious vulnerabilities in the OSS model, it also presents an opportunity for the software engineering community to innovate and adapt. By learning from this incident and implementing robust safeguards, we can work towards an open-source ecosystem that maintains its innovative spirit and collaborative nature while significantly enhancing its security and sustainability. 

As we have Digital Forensics and Incident Response tools for classical
security incidents, we also need similar solutions tailored to attacks
on software repositories. In sophisticated cases like the XZ Utils
attack, we need tools designed to reconstruct or clean up repository
artifacts post-attack. Here, MSR techniques can be used to both
proactively detect unusual project behavior and assist in
reconstructing attacks with data-driven evidence.

\section*{Data Availability}
The scripts and the data collected for this study are available in our replication package at the following address: 
\url{https://github.com/przymusp/XZ-Attack}

\bibliography{references}


\begin{thebibliography}{21}


\ifx \showCODEN    \undefined \def \showCODEN     #1{\unskip}     \fi
\ifx \showDOI      \undefined \def \showDOI       #1{#1}\fi
\ifx \showISBNx    \undefined \def \showISBNx     #1{\unskip}     \fi
\ifx \showISBNxiii \undefined \def \showISBNxiii  #1{\unskip}     \fi
\ifx \showISSN     \undefined \def \showISSN      #1{\unskip}     \fi
\ifx \showLCCN     \undefined \def \showLCCN      #1{\unskip}     \fi
\ifx \shownote     \undefined \def \shownote      #1{#1}          \fi
\ifx \showarticletitle \undefined \def \showarticletitle #1{#1}   \fi
\ifx \showURL      \undefined \def \showURL       {\relax}        \fi
\providecommand\bibfield[2]{#2}
\providecommand\bibinfo[2]{#2}
\providecommand\natexlab[1]{#1}
\providecommand\showeprint[2][]{arXiv:#2}

\bibitem[gha(2011)]%
        {gharchive}
 \bibinfo{year}{2011}\natexlab{}.
\newblock \bibinfo{title}{{G}{H} {A}rchive}.
\newblock \bibinfo{howpublished}{\url{https://www.gharchive.org/}}.
\newblock
\newblock
\shownote{[Accessed 01-07-2024]}.


\bibitem[Canfora et~al\mbox{.}(2020)]%
        {canfora2020investigating}
\bibfield{author}{\bibinfo{person}{Gerardo Canfora}, \bibinfo{person}{Andrea
  Di~Sorbo}, \bibinfo{person}{Sara Forootani}, \bibinfo{person}{Antonio
  Pirozzi}, {and} \bibinfo{person}{Corrado~Aaron Visaggio}.}
  \bibinfo{year}{2020}\natexlab{}.
\newblock \showarticletitle{Investigating the vulnerability fixing process in
  OSS projects: Peculiarities and challenges}.
\newblock \bibinfo{journal}{\emph{Computers \& Security}}  \bibinfo{volume}{99}
  (\bibinfo{year}{2020}), \bibinfo{pages}{102067}.
\newblock


\bibitem[Coelho and Valente(2017)]%
        {10.1145/3106237.3106246}
\bibfield{author}{\bibinfo{person}{Jailton Coelho} {and}
  \bibinfo{person}{Marco~Tulio Valente}.} \bibinfo{year}{2017}\natexlab{}.
\newblock \showarticletitle{Why modern open source projects fail}. In
  \bibinfo{booktitle}{\emph{Proceedings of the 2017 11th Joint Meeting on
  Foundations of Software Engineering}} (Paderborn, Germany)
  \emph{(\bibinfo{series}{ESEC/FSE 2017})}. \bibinfo{publisher}{Association for
  Computing Machinery}, \bibinfo{address}{New York, NY, USA},
  \bibinfo{pages}{186–196}.
\newblock
\showISBNx{9781450351058}
\urldef\tempurl%
\url{https://doi.org/10.1145/3106237.3106246}
\showDOI{\tempurl}


\bibitem[Dias et~al\mbox{.}(2021)]%
        {dias2021makes}
\bibfield{author}{\bibinfo{person}{Edson Dias}, \bibinfo{person}{Paulo
  Meirelles}, \bibinfo{person}{Fernando Castor}, \bibinfo{person}{Igor
  Steinmacher}, \bibinfo{person}{Igor Wiese}, {and} \bibinfo{person}{Gustavo
  Pinto}.} \bibinfo{year}{2021}\natexlab{}.
\newblock \showarticletitle{What makes a great maintainer of open source
  projects?}. In \bibinfo{booktitle}{\emph{2021 IEEE/ACM 43rd International
  Conference on Software Engineering (ICSE)}}. IEEE, \bibinfo{pages}{982--994}.
\newblock


\bibitem[Dias et~al\mbox{.}(2018)]%
        {dias2018drives}
\bibfield{author}{\bibinfo{person}{Luis~Felipe Dias}, \bibinfo{person}{Igor
  Steinmacher}, {and} \bibinfo{person}{Gustavo Pinto}.}
  \bibinfo{year}{2018}\natexlab{}.
\newblock \showarticletitle{Who drives company-owned OSS projects: internal or
  external members?}
\newblock \bibinfo{journal}{\emph{Journal of the Brazilian Computer Society}}
  \bibinfo{volume}{24}, \bibinfo{number}{1} (\bibinfo{year}{2018}),
  \bibinfo{pages}{16}.
\newblock


\bibitem[Gasser et~al\mbox{.}(2014)]%
        {6838249}
\bibfield{author}{\bibinfo{person}{Oliver Gasser}, \bibinfo{person}{Ralph
  Holz}, {and} \bibinfo{person}{Georg Carle}.} \bibinfo{year}{2014}\natexlab{}.
\newblock \showarticletitle{A deeper understanding of SSH: Results from
  Internet-wide scans}. In \bibinfo{booktitle}{\emph{2014 IEEE Network
  Operations and Management Symposium (NOMS)}}. \bibinfo{pages}{1--9}.
\newblock
\urldef\tempurl%
\url{https://doi.org/10.1109/NOMS.2014.6838249}
\showDOI{\tempurl}


\bibitem[Guizani et~al\mbox{.}(2021)]%
        {guizani2021long}
\bibfield{author}{\bibinfo{person}{Mariam Guizani}, \bibinfo{person}{Amreeta
  Chatterjee}, \bibinfo{person}{Bianca Trinkenreich},
  \bibinfo{person}{Mary~Evelyn May}, \bibinfo{person}{Geraldine~J Noa-Guevara},
  \bibinfo{person}{Liam~James Russell}, \bibinfo{person}{Griselda~G
  Cuevas~Zambrano}, \bibinfo{person}{Daniel Izquierdo-Cortazar},
  \bibinfo{person}{Igor Steinmacher}, \bibinfo{person}{Marco~A Gerosa},
  {et~al\mbox{.}}} \bibinfo{year}{2021}\natexlab{}.
\newblock \showarticletitle{The long road ahead: Ongoing challenges in
  contributing to large oss organizations and what to do}.
\newblock \bibinfo{journal}{\emph{Proceedings of the ACM on Human-Computer
  Interaction}} \bibinfo{volume}{5}, \bibinfo{number}{CSCW2}
  (\bibinfo{year}{2021}), \bibinfo{pages}{1--30}.
\newblock


\bibitem[Hissam et~al\mbox{.}(2002)]%
        {hissam2002trust}
\bibfield{author}{\bibinfo{person}{Scott~A. Hissam}, \bibinfo{person}{Daniel
  Plakosh}, {and} \bibinfo{person}{C Weinstock}.}
  \bibinfo{year}{2002}\natexlab{}.
\newblock \showarticletitle{Trust and vulnerability in open source software}.
\newblock \bibinfo{journal}{\emph{IEE Proceedings-Software}}
  \bibinfo{volume}{149}, \bibinfo{number}{1} (\bibinfo{year}{2002}),
  \bibinfo{pages}{47--51}.
\newblock


\bibitem[Hunt(2019)]%
        {hunt2019have}
\bibfield{author}{\bibinfo{person}{Troy Hunt}.}
  \bibinfo{year}{2019}\natexlab{}.
\newblock \showarticletitle{Have i been pwned}.
\newblock \bibinfo{journal}{\emph{URL: https://haveibeenpwned. com (visited on
  10/15/2020)}} (\bibinfo{year}{2019}).
\newblock


\bibitem[Lenarduzzi et~al\mbox{.}(2020)]%
        {lenarduzzi2020open}
\bibfield{author}{\bibinfo{person}{Valentina Lenarduzzi},
  \bibinfo{person}{Davide Taibi}, \bibinfo{person}{Davide Tosi},
  \bibinfo{person}{Luigi Lavazza}, {and} \bibinfo{person}{Sandro Morasca}.}
  \bibinfo{year}{2020}\natexlab{}.
\newblock \showarticletitle{Open source software evaluation, selection, and
  adoption: a systematic literature review}. In \bibinfo{booktitle}{\emph{2020
  46th Euromicro Conference on Software Engineering and Advanced Applications
  (SEAA)}}. IEEE, \bibinfo{pages}{437--444}.
\newblock


\bibitem[Lifshitz-Assaf and Nagle(2021)]%
        {lifshitz2021digital}
\bibfield{author}{\bibinfo{person}{Hila Lifshitz-Assaf} {and}
  \bibinfo{person}{Frank Nagle}.} \bibinfo{year}{2021}\natexlab{}.
\newblock \showarticletitle{The digital economy runs on open source. Here's how
  to protect it}.
\newblock \bibinfo{journal}{\emph{Harvard Business Review}}
  \bibinfo{volume}{2} (\bibinfo{year}{2021}).
\newblock


\bibitem[Mouratidis et~al\mbox{.}(2005)]%
        {mouratidis2005security}
\bibfield{author}{\bibinfo{person}{Haralambos Mouratidis},
  \bibinfo{person}{Paolo Giorgini}, {and} \bibinfo{person}{Gordon Manson}.}
  \bibinfo{year}{2005}\natexlab{}.
\newblock \showarticletitle{When security meets software engineering: a case of
  modelling secure information systems}.
\newblock \bibinfo{journal}{\emph{Information Systems}} \bibinfo{volume}{30},
  \bibinfo{number}{8} (\bibinfo{year}{2005}), \bibinfo{pages}{609--629}.
\newblock


\bibitem[O'Neill({[n.\,d.]})]%
        {oneillInternetRunsFree}
\bibfield{author}{\bibinfo{person}{Patrick~Howell O'Neill}.}
  \bibinfo{year}{[n.\,d.]}\natexlab{}.
\newblock \bibinfo{title}{The Internet Runs on Free Open-Source Software.
  {{Who}} Pays to Fix It?}
\newblock
  \bibinfo{howpublished}{https://www.technologyreview.com/2021/12/17/1042692/log4j-internet-open-source-hacking/}.
\newblock


\bibitem[Payne(2002)]%
        {payne2002security}
\bibfield{author}{\bibinfo{person}{Christian Payne}.}
  \bibinfo{year}{2002}\natexlab{}.
\newblock \showarticletitle{On the security of open source software}.
\newblock \bibinfo{journal}{\emph{Information systems journal}}
  \bibinfo{volume}{12}, \bibinfo{number}{1} (\bibinfo{year}{2002}),
  \bibinfo{pages}{61--78}.
\newblock


\bibitem[{Piotr Przymus} et~al\mbox{.}(2024)]%
        {piotr_przymus_patchscope_2024}
\bibfield{author}{\bibinfo{person}{{Piotr Przymus}}, \bibinfo{person}{{Jakub
  Narębski}}, \bibinfo{person}{{Mikołaj Fejzer}}, {and}
  \bibinfo{person}{{Krzysztof Stencel}}.} \bibinfo{year}{2024}\natexlab{}.
\newblock \bibinfo{title}{{PatchScope} – {A} {Modular} {Tool} for
  {Annotating} and {Analyzing} {Contributions}}.
\newblock
\newblock
\urldef\tempurl%
\url{https://ncusi.github.io/PatchScope/articles/patchscope_2024.pdf}
\showURL{%
\tempurl}
\newblock
\shownote{(under review)}.


\bibitem[Register(2020)]%
        {corejs}
\bibfield{author}{\bibinfo{person}{The Register}.}
  \bibinfo{year}{2020}\natexlab{}.
\newblock \bibinfo{title}{{W}hat happens when the maintainer of a {J}{S}
  library downloaded 26m times a week goes to prison for killing someone with a
  motorbike? {C}ore-js just found out}.
\newblock
  \bibinfo{howpublished}{\url{https://www.theregister.com/2020/03/26/corejs_maintainer_jailed_code_release}}.
\newblock
\newblock
\shownote{[Accessed 28-06-2024]}.


\bibitem[Scalco et~al\mbox{.}(2022)]%
        {scalco2022feasibility}
\bibfield{author}{\bibinfo{person}{Simone Scalco}, \bibinfo{person}{Ranindya
  Paramitha}, \bibinfo{person}{Duc-Ly Vu}, {and} \bibinfo{person}{Fabio
  Massacci}.} \bibinfo{year}{2022}\natexlab{}.
\newblock \showarticletitle{On the feasibility of detecting injections in
  malicious npm packages}. In \bibinfo{booktitle}{\emph{Proceedings of the 17th
  International Conference on Availability, Reliability and Security}}.
  \bibinfo{pages}{1--8}.
\newblock


\bibitem[Tulili et~al\mbox{.}(2023)]%
        {tulili2023burnout}
\bibfield{author}{\bibinfo{person}{Tien~Rahayu Tulili}, \bibinfo{person}{Andrea
  Capiluppi}, {and} \bibinfo{person}{Ayushi Rastogi}.}
  \bibinfo{year}{2023}\natexlab{}.
\newblock \showarticletitle{Burnout in software engineering: A systematic
  mapping study}.
\newblock \bibinfo{journal}{\emph{Information and Software Technology}}
  \bibinfo{volume}{155} (\bibinfo{year}{2023}), \bibinfo{pages}{107116}.
\newblock


\bibitem[Wang et~al\mbox{.}(2019)]%
        {wang2019detecting}
\bibfield{author}{\bibinfo{person}{Xinda Wang}, \bibinfo{person}{Kun Sun},
  \bibinfo{person}{Archer Batcheller}, {and} \bibinfo{person}{Sushil Jajodia}.}
  \bibinfo{year}{2019}\natexlab{}.
\newblock \showarticletitle{Detecting" 0-day" vulnerability: An empirical study
  of secret security patch in OSS}. In \bibinfo{booktitle}{\emph{2019 49th
  Annual IEEE/IFIP International Conference on Dependable Systems and Networks
  (DSN)}}. IEEE, \bibinfo{pages}{485--492}.
\newblock


\bibitem[Zhou et~al\mbox{.}(2024)]%
        {zhou2024oss}
\bibfield{author}{\bibinfo{person}{Xiaoyan Zhou}, \bibinfo{person}{Ying Zhang},
  \bibinfo{person}{Wenjia Niu}, \bibinfo{person}{Jiqiang Liu},
  \bibinfo{person}{Haining Wang}, {and} \bibinfo{person}{Qiang Li}.}
  \bibinfo{year}{2024}\natexlab{}.
\newblock \showarticletitle{OSS Malicious Package Analysis in the Wild}.
\newblock \bibinfo{journal}{\emph{arXiv preprint arXiv:2404.04991}}
  (\bibinfo{year}{2024}).
\newblock


\bibitem[Zimmermann et~al\mbox{.}(2019)]%
        {zimmermann2019small}
\bibfield{author}{\bibinfo{person}{Markus Zimmermann},
  \bibinfo{person}{Cristian-Alexandru Staicu}, \bibinfo{person}{Cam Tenny},
  {and} \bibinfo{person}{Michael Pradel}.} \bibinfo{year}{2019}\natexlab{}.
\newblock \showarticletitle{Small world with high risks: A study of security
  threats in the npm ecosystem}. In \bibinfo{booktitle}{\emph{28th USENIX
  Security symposium (USENIX security 19)}}. \bibinfo{pages}{995--1010}.
\newblock


\end{thebibliography}

\end{document}